\documentclass{article}

\PassOptionsToPackage{numbers, compress}{natbib}

\usepackage{tcolorbox}
\usepackage{tikz}
\tcbuselibrary{skins,breakable}

\tcbset{
  enhanced,
  colback=gray!10,
  colframe=black!70,
  arc=3mm,
  boxrule=0.5pt,
  left=8pt,right=8pt,top=8pt,bottom=8pt,
  fonttitle=\bfseries,
  breakable
}

\newcommand{\dashline}{\vspace{4pt}\noindent\centerline{\rule[0.5ex]{0.9\linewidth}{0.4pt}}\vspace{6pt}}
\usepackage[final]{neurips_2023}
\usepackage{tcolorbox}
\usepackage{pifont} 
\usepackage{graphicx}
\usepackage{tabularx}
\usepackage{tabularx}
\usepackage{ragged2e}
\usepackage{booktabs}
\usepackage{tikz}
\usepackage{subcaption}
\usepackage{makecell}
\usepackage{amsmath}
\usepackage{pifont}

\newcolumntype{Y}{>{\raggedleft\arraybackslash}X} 
\usetikzlibrary{arrows.meta, positioning, shapes.multipart, calc}
\usetikzlibrary{positioning, arrows.meta}
\tikzset{
  block/.style = {draw, rectangle, minimum height=1.2cm, minimum width=3.5cm, align=center},
  agent/.style = {draw, rectangle, rounded corners, minimum height=1.1cm, minimum width=3.5cm, align=center},
  arrow/.style = {thick, ->, >=stealth},
  feedback/.style = {thick, dashed, ->, >=stealth},
}

\usepackage[utf8]{inputenc} 
\usepackage[T1]{fontenc}    
\usepackage{hyperref}       
\usepackage{url}            
\usepackage{booktabs}       
\usepackage{amsfonts}       
\usepackage{nicefrac}       
\usepackage{microtype}      
\usepackage{xcolor}         
\usepackage{multirow}
\usepackage{float}
\usepackage{natbib}

\usepackage{comment}
\usepackage{caption}

\title{Orchestration Framework for Financial Agents:\\ From Algorithmic Trading to Agentic Trading}

\author{
Jifeng Li$^{1}$,
Arnav Grover$^{2}$,
Abraham Alpuerto$^{3}$,
Yupeng Cao$^{4}$,
Xiao-Yang Liu$^{1}$\thanks{Corresponding author.} \\
\\[-2mm]
$^{1}$SecureFinAI Lab, Columbia University, 
$^{2}$Purdue University, \\
$^{3}$Rensselaer Polytechnic Institute, 
$^{4}$Stevens Institute of Technology
}

\begin{document}


\maketitle

\begin{abstract}

The financial market is a mission-critical playground for AI agents due to its temporal dynamics and low signal-to-noise ratio. Building an effective algorithmic trading system may require a professional team to develop and test over the years. In this paper, we propose an orchestration framework for financial agents, which aims to democratize financial intelligence to the general public. We map each component of the traditional algorithmic trading system to agents, including planner, orchestrator, alpha agents, risk agents, portfolio agents, backtest agents, execution agents, audit agents, and memory agent.
We present two in-house trading examples. 
For the stock trading task (hourly data from 04/2024 to 12/2024), our approach achieved a return of $20.42\%$, a Sharpe ratio of 2.63, and a maximum drawdown of $-3.59\%$, while the S\&P 500 index yielded a return of $15.97\%$. 
For the BTC trading task (minute data from 27/07/2025 to 13/08/2025), our approach achieved a return of $8.39\%$, a Sharpe ratio of $0.38$, and a maximum drawdown of $-2.80\%$, whereas the BTC price increased by $3.80\%$. Our code is available on \href{https://github.com/Open-Finance-Lab/AgenticTrading}{GitHub}.

\end{abstract}
\vspace{-0.1in}

\vspace{-0.1in}
\section{Introduction}

From floor trading with chalkboards and open outcry to telephone order routing, and then to algorithmic trading, the market microstructure has reorganized how orders are created, conveyed, and executed \cite{treleaven2013algorithmic,aldridge2013hft,kissell2013science}. The algorithmic trading (AT) system \cite{treleaven2013algorithmic} follows a pipeline from processing financial data, extracting trading signals, portfolio management to execution and evaluation. Designing an effective AT system may require a professional team to develop and test over years. Recent works have demonstrated the great potential of AI agents: reasoning-and-acting \cite{yao2023react}, self-teaching for using tools \cite{schick2023toolformer}, generative agents \cite{park2023generative}, reflection and memory \cite{shinn2023reflexion}, and multi-agent role coordination \cite{li2023camel}. The financial market is a particularly challenging playground for AI agents due to its unique features of temporal dynamics and low signal-to-noise ratio. In particular, \textbf{agentic trading} is a mission-critical task in a high-stakes domain.



In this paper, we propose an end-to-end orchestration framework for financial agents, which maps the components of the traditional AT system to agents and democratizes financial intelligence to the general public. First, we map each component of the AT system to agents, including planner, orchestrator, alpha agents, risk agents, portfolio agents, backtest agents, execution agents, audit agents, and memory agent. Second, we use the Model Context Protocol (MCP) for control messages between the orchestrator and agents and the Agent-to-Agent protocol (A2A) for communication among agents, while a memory agent records states, prompts, tool calls, and decisions. 

Finally, we develop two homegrown trading examples. For the stock trading task backtested from 04/2024 to 01/2025 (hourly data), our agents achieve a return of $20.42\%$, volatility of $11.83\%$ and Sharpe ratio of $2.63$ with max drawdown of $-3.59\%$, while the S\&P 500 index has a return of $15.97\%$; however, the equally weighted method with weekly rebalance has a return of $47.46\%$. For the BTC trading task backtested from 27/07/2025 to 13/08/2025 (minute data), our agents achieve a return of $8.39\%$, volatility of $24.23\%$ and Sharpe ratio of $0.378$ with max drawdown of $-2.80\%$, while the BTC price increased $3.80\%$. 

\begin{figure}[t]
    \centering
    \includegraphics[width=1\linewidth]{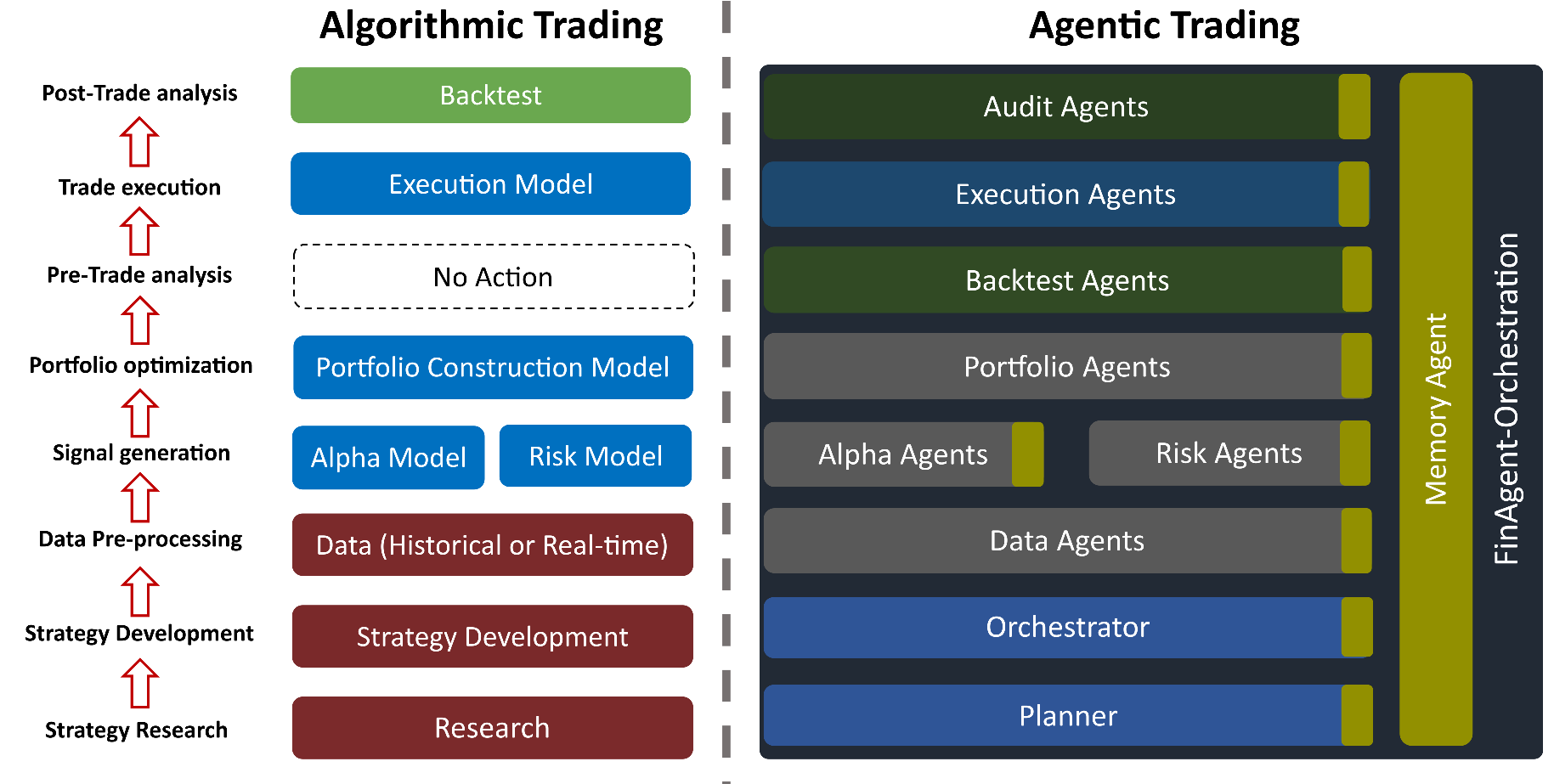}
    \caption{Agentic trading vs. algorithmic trading: we map the AT components to agents in our FinAgent orchestration framework, where a memory agent provides the contexts to other agents.
    }
    \label{fig:arch_compare}
    \vspace{-0.2in}
\end{figure}

\vspace{-0.1in}
\section{Proposed Framework for FinAgent Orchestration}
\vspace{-0.1in}

\subsection{Overview}
\vspace{-0.1in}

We build a FinAgent orchestration framework structured around multiple agent pools, each orchestrating a stage (data, alpha, risk, portfolio, execution) so that the system runs end-to-end from raw data to trading orders. We use several LLM models (e.g., GPT-4o \cite{hello_gpt4o,gpt4o_system_card}, Llama3 \cite{dubey2024llama}, FinGPT \cite{liu2023fingpt}) to power different agents. Data agents pull from multiple sources (e.g., Polygon and yfinance) \cite{polygon_api,yfinance_api}. We compare coverage, consistency, and delay across sources and keep the better ones; we then align time and symbols, clean errors and gaps, and form simple features \cite{liu2022finrlmeta, yang2020qlib}.
The cleaned data is fed into alpha and risk agents. Alpha agents propose signal structures, while tool-based modules compute the numerical signals, and risk agents compute exposures and limits \cite{treleaven2013algorithmic,aldridge2013hft,kissell2013science}. We check signals with different metrics (e.g., rank-IC), rolling tests, and a walk-forward backtest \cite{liu2022finrlmeta, kissell2013science, yang2020qlib}.Signal diagnostics (e.g., rank-IC) are computed by tool modules and never exposed to LLMs.
Approved signals are fed into portfolio agents. 

We backtest long-only and long-short rules under capital and turnover constraints \cite{kissell2013science,aldridge2013hft}. The system keeps only signals, risk rules, and portfolios that pass these checks \cite{tradingagents,rdagents}.
Next, we run simulated or live trading and adjust settings by market. Evaluation and attribution agents verify equity curves, drawdowns, contributions, and write a full log \cite{tradingagents,rdagents}.
The planner and the orchestrator use these logs to update the plan for the next loop, while avoiding any use of evaluation-window outcomes in agent prompts. And the memory module keeps the state for reuse \cite{finmem,yu2024fincon,zhang2023marl}.
\vspace{-0.1in}

\subsection{Control Messages and Agents' Communications}
\vspace{-0.1in}

\textbf{Control Messages}.
All agent pools are controlled by the orchestrator through MCP. The orchestrator sends small control messages that describe the task (node type, task id, declared inputs with schemas, policy flags, timeout, retry budget) and waits for a reply (acknowledgement, status, logs, artifact ids); it also tracks health with heartbeats and monitors completion until the tasks finish \cite{mcp_website,mcp_repo,tradingagents,rdagents}. MCP exposes each agent pool as a tool-like endpoint with a unified request–response schema.\cite{mcp_website,mcp_repo}. Inside a pool, a manager agent breaks a task into subtasks and assigns them to subordinate agents for cooperative execution; partial results are merged and returned upstream \cite{li2023camel,wu2024autogen}.

\textbf{Agents' communications}.
After a task is issued, each agent pool uses A2A to talk with the memory agent to read prior context and to upload logs and key results, so that state and rationale persist across runs \cite{finmem,yu2024fincon,tradingagents,rdagents}. Memory stores only structural summaries, not evaluation-window labels. Within a pool, agents of the same or different types also coordinate over A2A to complete the task; messages use simple types (ask, tell, propose, confirm) with role tags and context ids, following standard agent communication patterns \cite{fipa_acl,zhu2022comm_marl}. Agents share progress at fixed intervals; if a collaborator fails, a peer can take over or the orchestrator can reassign the job. All peer exchanges are time-stamped and stored in memory for replay and audit \cite{finmem,yu2024fincon}.
\vspace{-0.1in}


\vspace{-0.1in}
\section{Homegrown Trading Examples}
\vspace{-0.1in}

\subsection{Stock Trading and Crypto Trading Tasks}
\vspace{-0.1in}

\textbf{Stock backtesting pipeline.}  
We employ a unified plan graph and interface; only data sources, horizons, and scheduling change. 
For stocks, data agents fetch hourly bars from Polygon and yfinance, dedupe and align calendars, 
correct anomalies, and compute baseline features (returns, momentum, volatility, volume ratios) 
\cite{polygon_api,yfinance_api,liu2022finrlmeta,yang2020qlib}. 
Following our prompt-design constraints, the Alpha Agents propose factor structures based only 
on published literature and do not access any evaluation-window data. 
All numerical signal construction and return mapping are handled by tool-based modules. 
Risk agents compute exposures (market, sector, name) and enforce constraints 
(concentration, volatility, drawdown). 
Portfolio agents test long-only and long–short rules under capital and turnover limits; 
execution translates weights to orders with slippage/transaction cost models and reconciles fills. 
Evaluation runs walk-forward backtesting, attribution, and metric aggregation using the same plan graph, 
while keeping realized returns and performance metrics hidden from LLM agents 
\cite{treleaven2013algorithmic,aldridge2013hft,kissell2013science}.
\textbf{Crypto backtesting pipeline.}  
The BTC pipeline reuses the same plan graph and message schema, but with minute-level bars and
intraday prompts. Data agents ingest minute bars from Polygon \cite{polygon_api}, dedupe and align
timestamps, and aggregate features on a decision clock. Following the same prompt-design rules,
the Alpha Agents propose short-horizon microstructure factor structures (order flow imbalance,
spread, volume spikes) based only on prior literature and do not access any evaluation-window data.
All numerical feature transformations and signal computations are carried out by tool-based modules. 
Risk agents impose stricter caps on realized volatility, position size, and drawdown, and apply drift
or volatility gates before execution. Portfolio sizing follows a long-flat regime with turnover control.
Execution sends orders only when all gates pass and logs fills considering latency. Evaluation 
aggregates results to daily metrics (volatility, Sharpe, drawdown) while keeping realized returns 
and performance outcomes hidden from LLM agents, maintaining the same data path as the stock 
pipeline \cite{liu2022finrlmeta,tradingagents,rdagents}.

\vspace{-0.1in}
\subsection{Backtesting Performance}
\vspace{-0.1in}

\textbf{Experimental setup}. 
We run one orchestration pipeline across stocks and BTC. The initial assets are all \$100,000, and transactions are conducted in units of total dollar amount. The plan graph is similar, while schedules and rebalancing methods differ by markets. For stocks we backtest on a static seven-stock universe (AAPL, MSFT, GOOGL, JPM, TSLA, NVDA, META) from 09/2022 to 01/2025 by hourly. We use GPT-4o \cite{hello_gpt4o,gpt4o_system_card} to survey prior factor studies and to draft feature lists; in all prompts we restrict materials to published sources and do not expose any test-period market data, so the LLM cannot leak labels or prices from the backtest window. Baselines are the S\&P 500 (SPY), QQQ, IWM, VTI, and an equal-weighted portfolio (weekly rebalance). For BTC we use minute bars from 05/2025 to 08/2025; positions update every minute, and trades trigger only when drift and rule gates are met; the baseline is Buy\&Hold.
\begin{table}[t]
\centering
\footnotesize
\setlength{\tabcolsep}{3pt}
\renewcommand{\arraystretch}{1.1}
\caption{Comparison of trading performance. Arrows indicate: \(\uparrow\)=higher is better; \(\downarrow\)=lower is better (for Max Drawdown, less negative is better). 
B\&H denotes a Buy \& Hold strategy, where other ETFs are purchased at the start and held for the entire evaluation period. 
EW refers to an equally weighted portfolio constructed across all selected stocks or assets. 
Sharpe ratios are computed from daily/weekly returns with $R_f = 0$ due to short evaluation horizons. 
MDD: maximum drawdown.}
\label{tab:bench_comp_ours}
\begin{tabularx}{\linewidth}{l| *{6}{Y}| YY}
\toprule
Metric & Ours & SPY & QQQ & IWM & VTI & EW & \makecell{BTC\\(Ours)} & \makecell{BTC\\(B\&H)} \\
\midrule
Total Return $\uparrow$   & 20.42\% & 16.60\% & 21.59\% & 11.45\% & 16.29\% & \textbf{47.46\%} & \textbf{8.39\%} & 3.80\% \\
Annual Return $\uparrow$  & 31.08\% & 25.07\% & 32.94\% & 17.10\% & 24.59\% & \textbf{76.07\%} & -- & -- \\
Volatility $\downarrow$   & \textbf{11.83\%} & 13.49\% & 18.38\% & 21.61\% & 13.72\% & 22.54\% & \textbf{24.23}\% & 25.82\% \\
Sharpe Ratio $\uparrow$   & 2.63 & 1.86 & 1.79 & 0.79 & 1.79 & \textbf{3.37} & \textbf{0.378} & 0.170 \\
MDD (\%) $\uparrow$     & \textbf{-3.59} & -8.89 & -14.13 & -11.60 & -9.06 & -16.21 & \textbf{-2.80} & -5.26 \\
\bottomrule
\end{tabularx}
\vspace{-0.1in}
\end{table}

\begin{figure*}[t]
  \centering
  \includegraphics[width=\linewidth]{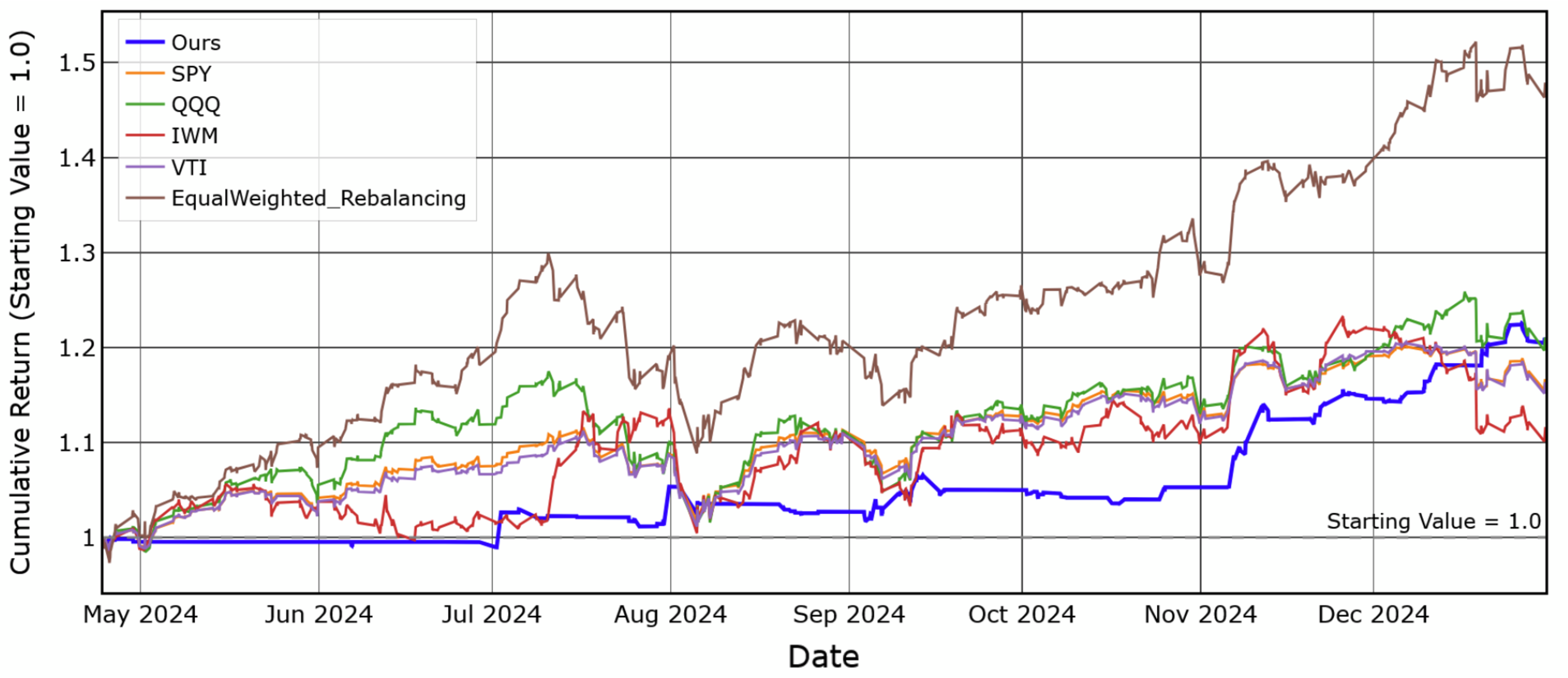}
   \caption{Seven-stock cumulative returns with the test window from 24/04/2024 to 31/12/2024 (within the 2022--2024 sample, the scrolling training window size is 3 months). The agentic strategy shows lower volatility and a smaller max drawdown, while the equally-weighted benchmark attains the highest total return. ETF baselines: SPY, QQQ, IWM, VTI. Metrics are reported in Table \ref{tab:bench_comp_ours}.}
  \label{fig:equity_perf}
  \vspace{-0.15in}
\end{figure*}

\begin{figure*}[t]
  \centering
  \includegraphics[width=\linewidth]{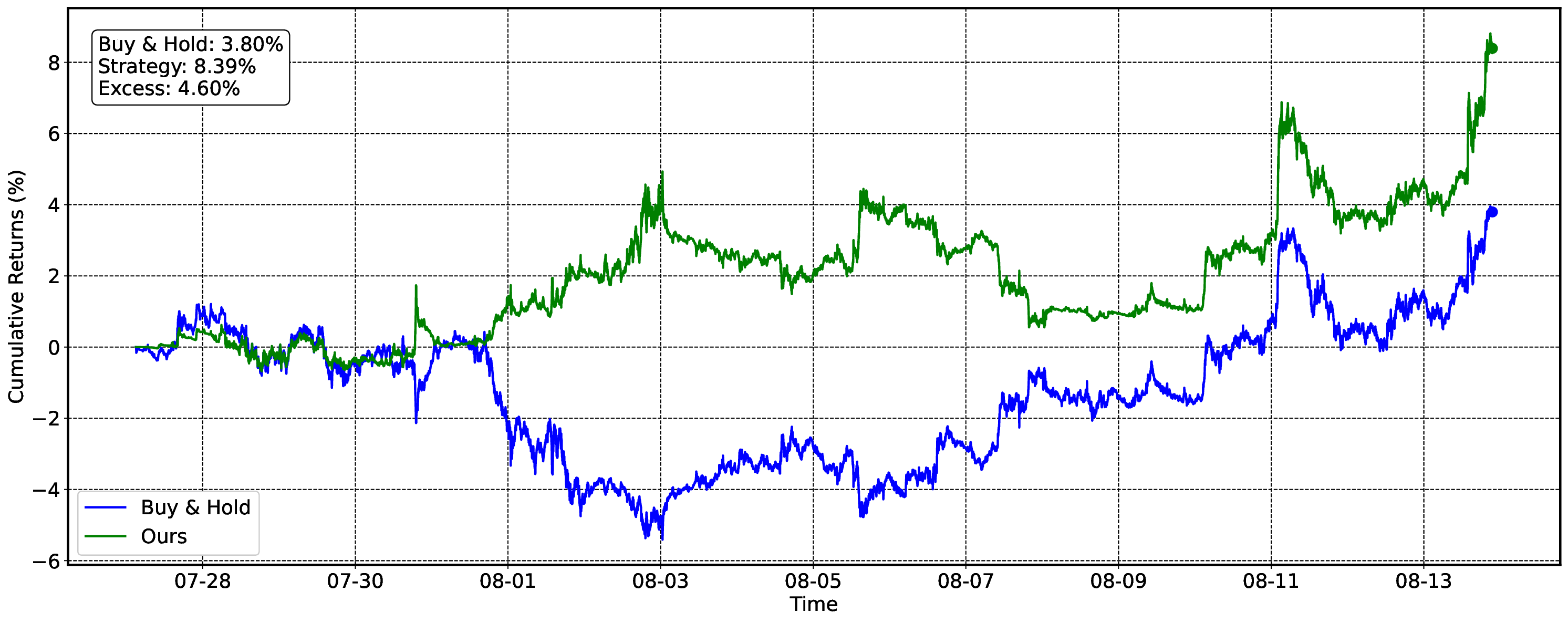}
  \caption{BTC results (07/27 to 08/13 in 2025, the scrolling window is 7 days). Cumulative returns: Buy\&Hold \(+3.80\%\), Ours \(+8.39\%\), Excess \(+4.59\%\). Excess \(=\) Ours \(-\) Buy\&Hold.}
  \label{fig:btc_perf}
  \vspace{-0.25in}
\end{figure*}
\textbf{Stocks (multi\textendash stock agentic portfolio)}.
Table~\ref{tab:bench_comp_ours} and Fig.~\ref{fig:equity_perf} report that the equally weighted benchmark attains the highest total return (47.46\%, Sharpe ratio 3.37). Ours delivers total return 20.42\%, the lowest volatility (11.83\%) and the smallest max drawdown ($-3.59\%$), with Sharpe ratio 2.63 (ETF range $0.79$--$1.86$). The profile is risk\textendash controlled and consistent with execution gated by the risk and portfolio pools.


\textbf{Cryptocurrency (BTC, minute data)}.
Fig.~\ref{fig:btc_perf} and Table~\ref{tab:bench_comp_ours} show that, on BTC/USDT,
the Buy\&Hold benchmark ends at +3.80\% while our strategy finishes at
approximately +8.4\%. Over this short 17-day window, the BTC strategy earns about 4.6 percentage points of excess return over Buy-and-Hold, with lower realized volatility and smaller max drawdown. This mirrors the equity experiment, where the agentic strategy trades off some total return for tighter risk. (\( \mathrm{Excess} = \mathrm{Ours} - \mathrm{Buy\&Hold} \)).

\vspace{-0.1in}
\section{Conclusion}
\vspace{-0.1in}

We proposed an orchestration framework for FinAgents, a paradigm shift from traditional AT system \cite{treleaven2013algorithmic,aldridge2013hft,kissell2013science} toward agentic trading. We map the components of the traditional AT system into agents, with standard agent protocols and a shared memory record for auditability \cite{mcp_website,zhu2022comm_marl,liu2022finrlmeta,zhang2023marl,yu2024fincon,finmem,tradingagents,rdagents}. Using one unified pipeline for stock and BTC tradings, we show a total return over the S\&P 500 index and ETFs; on BTC minute data, ours is +8.39\% versus Buy\&Hold +3.80\%. 
 
 Future work includes longer horizons and markets, ablations on gating/memory/messaging, adaptive planner updates under regime shifts \cite{guo2017quant}, broader information sources for signals \cite{ding2015deep,nassirtoussi2014text}, and releasing benchmarks and logs for replication.



\clearpage

\bibliographystyle{plainnat}
\bibliography{custom}

@article{treleaven2013algorithmic,
  title={Algorithmic trading review},
  author={Treleaven, Philip and Galas, Michal and Lalchand, Vidhi},
  journal={Communications of the ACM},
  volume={56},
  number={11},
  pages={76--85},
  year={2013},
  publisher={ACM}
}

@inproceedings{liu2022finrlmeta,
  title={{FinRL-Meta}: A Universe of Near-Real Market Environments for Data-Driven Deep Reinforcement Learning in Quantitative Finance},
  author={Liu, Xiao-Yang and Rui, Jingyang and Gao, Jiechao and Yang, Liuqing and Yang, Hongyang and Wang, Christina Dan and Guo, Jian and Wang, Zhaoran},
  booktitle={NeurIPS 2021 Workshop on Data-Centric AI},
  year={2022}
}

@misc{polygon_api,
  title={Polygon.io API Documentation},
  howpublished={\url{https://polygon.io/docs/stocks/getting-started}},
  note={Accessed: 2025-08-11}
}

@misc{yfinance_api,
  title={yfinance Python Library},
  author={Taha, Ran and others},
  howpublished={\url{https://pypi.org/project/yfinance/}},
  note={Accessed: 2025-08-11}
}

@inproceedings{zhang2023marl,
  title={Optimizing Trading Strategies in Quantitative Markets using Multi-Agent Reinforcement Learning},
  author={Zhang, Hengxi and Shi, Zhendong and Hu, Yuanquan and Ding, Wenbo and Kuruoglu, Ercan E. and Zhang, Xiao-Ping},
  booktitle={arXiv preprint arXiv:2303.11959},
  year={2023}
}

@article{ding2015deep,
  title={Deep learning for event-driven stock prediction},
  author={Ding, Xiao and Zhang, Yue and Liu, Ting and Duan, Junwen},
  journal={AAAI},
  year={2015}
}

@article{nassirtoussi2014text,
  title={Text mining for market prediction: A systematic review},
  author={Nassirtoussi, Amir H. and Aghabozorgi, Saeed and Wah, Teh Ying and Ngo, David C.L.},
  journal={Expert Systems with Applications},
  volume={41},
  number={16},
  year={2014}
}

@book{aldridge2013hft,
  title     = {High-Frequency Trading: A Practical Guide to Algorithmic Strategies and Trading Systems},
  author    = {Aldridge, Irene},
  year      = {2013},
  publisher = {Wiley},
  address   = {Hoboken, NJ}
}

@book{kissell2013science,
  title     = {The Science of Algorithmic Trading and Portfolio Management},
  author    = {Kissell, Robert},
  year      = {2013},
  publisher = {Academic Press},
  address   = {San Diego, CA}
}

@book{guo2017quant,
  author    = {Xin Guo and Tze Leung Lai and Howard Shek and Samuel Po{-}Shing Wong},
  title     = {Quantitative Trading: Algorithms, Analytics, Data, Models, Optimization},
  year      = {2017},
  publisher = {Chapman and Hall/CRC},
}

@article{yu2024fincon,
  title={Fincon: A synthesized llm multi-agent system with conceptual verbal reinforcement for enhanced financial decision making},
  author={Yu, Yangyang and Yao, Zhiyuan and Li, Haohang and Deng, Zhiyang and Jiang, Yuechen and Cao, Yupeng and Chen, Zhi and Suchow, Jordan and Cui, Zhenyu and Liu, Rong and others},
  journal={Advances in Neural Information Processing Systems},
  volume={37},
  pages={137010--137045},
  year={2024}
}

@article{finmem,
  title={Finmem: A performance-enhanced llm trading agent with layered memory and character design},
  author={Yu, Yangyang and Li, Haohang and Chen, Zhi and Jiang, Yuechen and Li, Yang and Suchow, Jordan W and Zhang, Denghui and Khashanah, Khaldoun},
  journal={IEEE Transactions on Big Data},
  year={2025},
  publisher={IEEE}
}

@article{tradingagents,
  title   = {{TradingAgents}: Multi-Agents LLM Financial Trading Framework},
  author  = {Yijia Xiao and Edward Sun and Di Luo and Wei Wang},
  journal = {arXiv preprint arXiv:2412.20138},
  year    = {2025},
}

@article{rdagents,
  title   = {R\&D-Agent-Quant: A Multi-Agent Framework for Data-Centric Factors and Model Joint Optimization},
  author  = {Yuante Li and Xu Yang and Xiao Yang and Minrui Xu and Xisen Wang and Weiqing Liu and Jiang Bian},
  journal = {arXiv preprint arXiv:2505.15155},
  year    = {2025},
  url     = {https://arxiv.org/abs/2505.15155}
}

@misc{mcp_website,
  author       = {{Model Context Protocol Working Group}},
  title        = {Model Context Protocol (MCP): Official Documentation},
  howpublished = {\url{https://modelcontextprotocol.io/}},
  year         = {2025},
  note         = {Accessed: 2025-08-29}
}

@misc{mcp_repo,
  author       = {{Model Context Protocol Working Group}},
  title        = {Model Context Protocol (MCP): Specification and Documentation Repository},
  howpublished = {\url{https://github.com/modelcontextprotocol/modelcontextprotocol}},
  year         = {2025},
  note         = {Accessed: 2025-08-29}
}

@misc{fipa_acl,
  author       = {{FIPA}},
  title        = {FIPA Agent Communication Language (ACL): Message Structure Specification},
  howpublished = {\url{https://www.fipa.org/specs/fipa00061/SC00061G.html}},
  year         = {2002},
  note         = {Accessed: 2025-08-29}
}

@article{zhu2022comm_marl,
  author       = {Changxi Zhu and Mehdi Dastani and Shihan Wang},
  title        = {A Survey of Multi-Agent Deep Reinforcement Learning with Communication},
  journal      = {arXiv preprint arXiv:2203.08975},
  year         = {2022},
  url          = {https://arxiv.org/abs/2203.08975}
}

@inproceedings{yao2023react,
  title={React: Synergizing reasoning and acting in language models},
  author={Yao, Shunyu and Zhao, Jeffrey and Yu, Dian and Du, Nan and Shafran, Izhak and Narasimhan, Karthik and Cao, Yuan},
  booktitle={International Conference on Learning Representations (ICLR)},
  year={2023}
}

@article{schick2023toolformer,
  title={Toolformer: Language models can teach themselves to use tools},
  author={Schick, Timo and Dwivedi-Yu, Jane and Dess{\`\i}, Roberto and Raileanu, Roberta and Lomeli, Maria and Hambro, Eric and Zettlemoyer, Luke and Cancedda, Nicola and Scialom, Thomas},
  journal={Advances in Neural Information Processing Systems},
  volume={36},
  pages={68539--68551},
  year={2023}
}

@article{shinn2023reflexion,
  title={Reflexion: Language agents with verbal reinforcement learning},
  author={Shinn, Noah and Cassano, Federico and Gopinath, Ashwin and Narasimhan, Karthik and Yao, Shunyu},
  journal={Advances in Neural Information Processing Systems},
  volume={36},
  pages={8634--8652},
  year={2023}
}

@inproceedings{park2023generative,
  title={Generative agents: Interactive simulacra of human behavior},
  author={Park, Joon Sung and O'Brien, Joseph and Cai, Carrie Jun and Morris, Meredith Ringel and Liang, Percy and Bernstein, Michael S},
  booktitle={Proceedings of the 36th Annual ACM Symposium on User Interface Software and Technology},
  pages={1--22},
  year={2023}
}

@article{li2023camel,
  title={Camel: Communicative agents for" mind" exploration of large language model society},
  author={Li, Guohao and Hammoud, Hasan and Itani, Hani and Khizbullin, Dmitrii and Ghanem, Bernard},
  journal={Advances in Neural Information Processing Systems},
  volume={36},
  pages={51991--52008},
  year={2023}
}

@inproceedings{wu2024autogen,
  title={Autogen: Enabling next-gen LLM applications via multi-agent conversations},
  author={Wu, Qingyun and Bansal, Gagan and Zhang, Jieyu and Wu, Yiran and Li, Beibin and Zhu, Erkang and Jiang, Li and Zhang, Xiaoyun and Zhang, Shaokun and Liu, Jiale and others},
  booktitle={First Conference on Language Modeling},
  year={2024}
}

@article{yang2020qlib,
  title={Qlib: An ai-oriented quantitative investment platform},
  author={Yang, Xiao and Liu, Weiqing and Zhou, Dong and Bian, Jiang and Liu, Tie-Yan},
  journal={arXiv preprint arXiv:2009.11189},
  year={2020}
}

@online{hello_gpt4o,
  title   = {Hello GPT-4o},
  author  = {OpenAI},
  year    = {2024},
  url     = {https://openai.com/index/hello-gpt-4o/},
  note    = {Accessed 2025-09-01}
}

@online{gpt4o_system_card,
  title   = {GPT-4o System Card},
  author  = {OpenAI},
  year    = {2024},
  url     = {https://openai.com/index/gpt-4o-system-card/},
  note    = {Accessed 2025-09-01}
}

@article{liu2023fingpt,
  title={{FinGPT}: Democratizing internet-scale data for financial large language models},
  author={Liu, Xiao-Yang and Wang, Guoxuan and Yang, Hongyang and Zha, Daochen},
  journal={Workshop on Instruction Tuning and Instruction Following, NeurIPS},
  year={2023}
}

@article{dubey2024llama,
  title={The {Llama 3} herd of models},
  author={Dubey, Abhimanyu and Jauhri, Abhinav and Pandey, Abhinav and Kadian, Abhishek and Al-Dahle, Ahmad and Letman, Aiesha and Mathur, Akhil and Schelten, Alan and Yang, Amy and Fan, Angela and others},
  journal={arXiv e-prints},
  pages={arXiv--2407},
  year={2024}
}

@article{quantagent,
  author       = {Xiong, F. and others},
  title        = {QuantAgent: Seeking Holy Grail in Trading by Self-Improving Large Language Model},
  journal      = {arXiv preprint},
  eprint       = {2509.09995},
  archivePrefix= {arXiv},
  year         = {2025}
}

@misc{alphaarena,
  author       = {A., B.},
  title        = {alpha-arena: A Benchmark for AI Investing Abilities},
  howpublished = {\url{https://github.com/AmadeusGB/alpha-arena}},
  note         = {GitHub Repository},
  year         = {2024}
}

@article{contesttrade,
  author       = {Zhang, R. and others},
  title        = {ContestTrade: A Multi-Agent Trading System Based on Internal Contest Mechanism},
  journal      = {arXiv preprint},
  eprint       = {2508.00554},
  archivePrefix= {arXiv},
  year         = {2025}
}

@article{stockagent,
  author       = {Jin, M. and others},
  title        = {When AI Meets Finance (StockAgent): Large Language Model-based Stock Trading in Simulated Real-world Environments},
  journal      = {arXiv preprint},
  eprint       = {2407.18957},
  archivePrefix= {arXiv},
  year         = {2024}
}

@misc{aihedgefund_singh_2025,
  author       = {Virat Singh},
  title        = {AI Hedge Fund: An AI Hedge Fund Team},
  howpublished = {\url{https://github.com/virattt/ai-hedge-fund}},
  year         = {2025},
  note         = {GitHub repository, accessed 2025-11-16}
}
\clearpage
\appendix
\section{More Test Details}
\textbf{Crypto backtesting pipeline.}
The BTC experiment uses the same agent classes and DAG topology as the equity pipeline, but runs on minute-level intraday data. Data Agents load raw OHLCV bars, funding rates, and open interest from Polygon~\cite{polygon_api}, deduplicate timestamps, align to the exchange calendar, and build rolling microstructure features on a fixed decision clock (e.g., $k$-minute windows). All features are computed only from information available at or before each decision time; evaluation-window labels, summary statistics, and functions that depend on future data are not exposed to any agent.

Alpha Agents coordinate signal design. Given the feature schema and prior crypto microstructure literature, they specify short-horizon factor structures (e.g., order-flow imbalance, bid--ask spread and depth, volume or volatility spikes, funding-rate or open-interest signals) and issue tool calls to construct these factors numerically. Alpha Agents do not compute predictions and do not observe realized returns. Forecasting (directional or return-based) is performed by tool-based ML modules, such as Ridge or tree ensembles, trained on rolling in-sample windows with adjacent validation windows under chronological splits.

Risk Agents use tighter constraints than in the equity pipeline to reflect the higher volatility of BTC. They impose volatility targets, position-size caps, leverage limits, and drawdown thresholds. Additional drift and short-horizon volatility checks may block trades when intraday moves exceed predefined bounds. The Risk Agent outputs a single risk-adjusted exposure signal, which is passed to the Portfolio Agent.

Portfolio Agents map the risk-adjusted signal to long--flat target positions, apply smoothing to limit turnover, and enforce minimum-change thresholds to avoid small position updates. Execution Agents simulate intraday order placement using historical best bid/ask and depth snapshots, accounting for latency, fees, spreads, and partial fills. Orders are submitted only when all checks from Data, Alpha, Risk, and Portfolio have passed.


A Backtest Orchestrator runs this pipeline in a walk-forward setup, the same as in the equity case. Each block has a training window for ML tools, a validation window for hyperparameter and stability checks, and a separate test window used only for paper trading. Evaluation turns test-window trades into daily metrics such as realized volatility, Sharpe ratio, drawdown, and turnover~\cite{liu2022finrlmeta,tradingagents,rdagents}.
Realized returns and performance outcomes are never given to any LLM, matching the data-flow rules used in the equity pipeline.

\section{BTC/USDT Trading Strategy}
\label{app:btc_hft}

This appendix gives implementation details of the BTC/USDT high-frequency trading strategy used in our experiments. We summarize the feature design, the prediction model, the training procedure, and the basic safeguards against data leakage.

\subsection{Feature Engineering and Data Preprocessing}

We build a feature set with more than 100 inputs that describe price, volume, volatility, and trend at the 1-minute frequency. Basic price features include smoothed 5-minute returns, 15-minute realized volatility, 60-minute exponentially weighted moving (EWM) volatility, and the difference between the mid-price and the volume-weighted average price (VWAP). Technical indicators include relative strength index (RSI) with windows 14 and 30, MACD (line, signal, and histogram), and Bollinger Band features such as band position, band width, and simple overbought/oversold flags.

Momentum-related features cover multiple horizons $1/3/5/10/15/30/60/240$ minutes. They include price momentum, trend strength, trend consistency, trend alignment, breakout flags, price acceleration, momentum alignment, trend persistence, and a simple momentum quality score. Volatility features include proxies for volatility clustering (GARCH-style terms), volatility regime identifiers, within-bar price range, and range expansion signals. Additional features encode support and resistance levels, mean-reversion signals based on price $z$-scores, volume statistics, and price–volume divergence.

All raw features are smoothed with an exponentially weighted moving average with decay parameter $\alpha = 0.4$. We then apply a RobustScaler transformation to reduce the influence of outliers and heavy tails. Finally, we remove low-variance features using a variance threshold of $0.01$ and keep only the top 70\% of features ranked by model importance. This reduces the number of inputs and makes the model more stable.

\subsection{Model Architecture and Training Methodology}

We use an XGBoost regression model to predict the next-minute return $r_{t+1}$ from a vector of lagged features $\mathbf{x}_t$. The main hyperparameters are: 300 trees, maximum depth $6$, learning rate $0.08$, subsample rate $0.8$, column subsample rate $0.8$, $\ell_1$ regularization $0.01$, $\ell_2$ regularization $0.05$, minimum child weight $1$, split penalty (gamma) $0.0$, and up to 512 histogram bins.

Training follows a rolling walk-forward scheme. The model is retrained every 24 hours (1{,}440 minutes) using a minimum training window of 7 days (10{,}080 minutes), with a prediction horizon of 1 minute. At each step, historical data are split into a training window and a validation window. The validation window is used to monitor model performance and to adjust basic signal rules, but not to tune the strategy on realized test outcomes.

To avoid data leakage, all input features are computed using information up to time $t-1$, and the prediction target is the forward 1-minute return $r_{t+1}$, with at least a 2-minute gap between feature timestamps and labels. We use early stopping with a patience of 20 boosting rounds based on validation loss. At each retraining, we recompute feature importance on the training window and again keep only the top 70\% most informative features for the next model fit.

\subsection{Signal Generation and Market Regime Identification}

The trading signal combines two parts: a model prediction and simple price-based rules (price action). The price-action part uses the following components:
\begin{itemize}
    \item \textbf{Short-, medium-, and longer-horizon momentum:} a weighted sum of 1-, 5-, and 15-minute returns with weights 40\%, 40\%, and 20\%. We multiply this sum by 10 so that it has a similar scale as other components.
    \item \textbf{Mean reversion:} a weighted sum of 20- and 60-period price deviations from a recent average (e.g., $z$-scores), multiplied by 5. This term is large when price moves far away from its recent level.
    \item \textbf{Breakout:} a discrete signal with value $\pm 3.0$ when price breaks above or below recent highs or lows.
    \item \textbf{Trend following:} a term that combines agreement of trends across horizons (trend alignment) with trend strength, multiplied by 100 to reflect its importance.
    \item \textbf{Momentum acceleration:} the change in the 5-minute return from one step to the next, multiplied by 20.
\end{itemize}
We mix the model prediction with the price-action signal based on a simple measure of model signal quality $q_t$ (for example, the absolute value of a standardized predicted return). The weight on the model is
\[
w_{\text{model}}(q_t) =
\begin{cases}
0.10, & q_t < 0.05, \\
0.20, & 0.05 \le q_t < 0.10, \\
0.40, & q_t \ge 0.10,
\end{cases}
\]
and the price-action part gets weight $1 - w_{\text{model}}(q_t)$. When the model signal is weak, we mostly follow price action; when the model signal is stronger, we give it more weight. We also classify simple market regimes using trend and volatility measures:
\begin{itemize}
    \item \textbf{Strong trend:} trends across several horizons point in the same direction more than 75\% of the time, and the trend strength score is above 0.1\%.
    \item \textbf{Breakout:} price moves above or below a recent 20-period high or low.
    \item \textbf{Sideways:} trend strength is below 0.05\% and a ratio of short-horizon to long-horizon volatility is above 1.2, suggesting choppy but directionless movement.
    \item \textbf{High volatility:} short-term volatility is more than 1.5 times long-term volatility.
\end{itemize}
Each regime uses a different mix of signal components. In strong-trend regimes, we emphasize momentum and trend signals and scale them by 1.5. In breakout regimes, we emphasize breakout and momentum-acceleration signals and double the breakout term. In sideways regimes, we put more weight on mean reversion and momentum and multiply the mean-reversion term by 1.2. In all other cases, we use a fixed mix of 70\% momentum and 30\% trend.

\subsection{Position Sizing and Risk Management}

Position size depends on the market regime we detect. We use a base size factor of 1.8 in strong-trend regimes, 2.5 in breakout regimes, 0.7 in sideways regimes, and 0.8 in high-volatility regimes. When the momentum score is higher than 70\% of its past values, we increase this base size by up to 30\%. Before trading, we first center and scale the raw signals using median absolute deviation (MAD), which measures how far values are from the middle, and then apply the hyperbolic tangent (tanh) function to keep very large values in a reasonable range. Entry thresholds are different in each regime: we use approximate percentile levels of the signal (45th percentile for trend regimes, 50th for breakout, 35th for sideways, and 40th for mixed regimes). Position limits keep each single position between 3\% and 5\% of capital and limit total leverage to 4.0.

Risk control combines fixed limits with rules based on cumulative loss (drawdown). We reduce positions when realized drawdowns grow: if drawdown goes beyond 1\%, 2\%, and 3\%, we cut position sizes by 20\%, 30\%, and 50\%, respectively. In addition, we check drawdowns at each step: if drawdown is above 1.5\% or 2.5\%, we immediately cut 50\% or 75\% of the current position. Each trade has a volatility-based stop-loss centered at $-0.8\%$, scaled between 0.5 and 1.5 times this level depending on current volatility. A maximum drawdown limit of $-3.0\%$ closes all positions when it is reached. To reduce trading frequency and avoid reacting to very short-lived noise, we smooth the final trading signal in two steps with exponentially weighted moving averages with decay parameters $\alpha_1 = 0.25$ and $\alpha_2 = 0.15$. We also use a band of width 0.08 where small changes do not change the position, and we require a minimum holding time of 8 minutes before reversing or closing positions.

\subsection{Backtesting Results and Performance Evaluation}

We test the strategy over a 17-day window on BTC/USDT (about 23{,}500 one-minute observations). Over this period, the strategy reaches a cumulative return of 8.39\%, while a Buy-and-Hold benchmark gains 3.80\%, so the excess return is $+4.59$ percentage points. On common risk measures, the strategy also improves on the benchmark. It has a Sharpe ratio of 0.380 versus 0.168 for Buy-and-Hold and a Calmar ratio of 166.06 versus 23.30. The maximum drawdown is smaller: $-2.80\%$ for the strategy versus $-5.26\%$ for Buy-and-Hold. The annualized volatility is slightly lower at 24.23\% (Buy-and-Hold: 25.82\%).

Trading activity is moderate for a high-frequency setting: there are 17 trades in total, or about 1.04 trades per day. The average holding time per trade is 16.07 hours, and the median holding time is 0.65 hours. The win rate is 64.7\% (Buy-and-Hold: 58.8\%), and the average daily return is 0.48\% (Buy-and-Hold: 0.23\%). Over the 17-day test window, the strategy achieves 8.39\% cumulative return versus 3.80\% for Buy-and-Hold, with lower volatility and max drawdown.

\section{Agentic Trading Projects}
Table~\ref{tab:fin_agent_comparison_grouped} shows that several open-source agent-based trading systems have attracted notable developer interest. The two multi–agent frameworks TradingAgents and AI Hedge Fund have about $24{,}800$ and $42{,}300$ GitHub stars and $4{,}600$ and $7{,}500$ forks, compared with smaller projects such as QuantAgent (306 stars, 71 forks) and ContestTrade (465 stars, 124 forks). All six projects have recent commits in late 2025, so they are being maintained rather than left idle.

Projects with more community activity also tend to use more agent-style designs. TradingAgents, AI Hedge Fund, and ContestTrade all include multi–agent analysis in their trading types and expose orchestration as a clear module; five of the six repositories provide some form of persistent memory. These higher-usage projects also cover several markets (equities, crypto, and derivatives) and are released as reusable frameworks or applications rather than single-use examples.  Table~\ref{tab:fin_agent_comparison_grouped} shows that several open-source projects with higher community activity already adopt multi-agent designs and some form of persistent memory. 
\newcommand{\cmark}{\ding{51}}
\newcommand{\xmark}{\ding{55}}
\newcolumntype{L}{>{\RaggedRight\arraybackslash}X}

\begin{table*}[t] 
    \centering
    \caption{Comparison of trading agents. 
    \cmark \ means the attribute is present; \xmark \ means it is not.}
    \label{tab:fin_agent_comparison_grouped}
    

    \begin{tabularx}{\textwidth}{l L L L L L L}
        \toprule
        \textbf{Attribute} & 
        \textbf{\shortstack[l]{Quant \\ Agent \\ \protect\cite{quantagent}}} &
        \textbf{\shortstack[l]{Alpha \\ Arena \\ \protect\cite{alphaarena}}} & 
        \textbf{\shortstack[l]{Trading \\ Agents \\ \protect\cite{tradingagents}}} & 
        \textbf{\shortstack[l]{AI Hedge \\ Fund \\ \protect\cite{aihedgefund_singh_2025}}} & 
        \textbf{\shortstack[l]{Contest \\ Trade \\ \protect\cite{contesttrade}}} & 
        \textbf{\shortstack[l]{Stock \\ Agent \\ \protect\cite{stockagent}}} \\
        
        \midrule
        \multicolumn{7}{l}{\textbf{Repository Attributes}} \\
        \midrule
        
        GitHub Stars & 
        306 &
        549 & 
        24800 & 
        42300 & 
        465 & 
        402 \\
        
        GitHub Forks & 
        71 &
        126 & 
        4600 & 
        7500 & 
        124 & 
        89 \\
        
        Last Update & 
        11/09/2025 &
        10/20/2025 & 
        10/09/2025 & 
        10/11/2025 & 
        10/13/2025 & 
        11/02/2025 \\
        
        License & 
        MIT &
        MIT & 
        Apache-2.0 & 
        MIT & 
        Apache-2.0 & 
        MIT \\
        
        \midrule
        \multicolumn{7}{l}{\textbf{Project Specifications}} \\
        \midrule
        
        Markets & 
        Equities, Forex, Crypto, Commodities &
        Crypto (Bitcoin, Ethereum) & 
        Equities (Any Ticker) & 
        Equities (Any Ticker) & 
        Equities (CN, US Markets) & 
        Equities (Any Ticker) \\
        
        Trading Types & 
        Technical Analysis, Pattern Recognition, Trend Analysis &
        Long Trading, Paper Trading & 
        Multi-Agent Analysis, Fundamental, Technical, Sentiment & 
        Multi-Agent Analysis, Valuation, Risk Management & 
        Event-Driven, Multi-Agent Analysis, Factor Portfolio & 
        LLM-Based Trading Simulation, Real-World Events \\
        
        Agents & 
        4 &
        2-6 & 
        6 & 
        18 & 
        2+ & 
        1 \\
        
        Tech Stack & 
        Python 3.10, conda, LangGraph &
        Python, LangChain, Streamlit & 
        Python, LLM, ReAct & 
        Python, Poetry, TypeScript, Ollama & 
        Python, LLM, React Loop & 
        Python 3.9, conda \\

        Project Type &
        Framework &
        Benchmark &
        Framework &
        Application &
        Framework &
        Application \\
        
        \midrule
        \multicolumn{7}{l}{\textbf{Agent Capabilities}} \\
        \midrule
        
        Memory & 
        \cmark &
        \xmark & 
        \cmark & 
        \cmark & 
        \cmark & 
        \cmark \\
        
        Orchestration & 
        \cmark &
        \xmark & 
        \cmark & 
        \cmark & 
        \cmark & 
        \xmark \\
        
        \bottomrule
\end{tabularx}
\end{table*}

\clearpage
\section{Prompt Design, Context Protocols, and Memory Integration}
\begin{tcolorbox}
\textbf{Data Agent – Stocks}\\
\textbf{Prompt:}\\
You are the Data Agent for US equities. Your task is to fetch, clean, align,
and normalize training-window market data from approved providers (Polygon,
yfinance). You may not infer or reference any future information beyond the
requested end date. You may not compute returns on the evaluation window. 
Return only cleaned features, metadata, and data-quality diagnostics.

\medskip
\textbf{Context:}\\
role=DataAgent, assetType=equity, frequency="1h", lookbackDays=180

\medskip
\textbf{Input:}\\
{"symbols":["AAPL","MSFT","GOOGL",...],\\
 "sourceHints":["polygon","yfinance"],\\
 "requiredFields":["open","high","low","close","volume"]}

\medskip
\textbf{Output (JSON):}
\begin{verbatim}
{
  "cleanedDataRef": "s3://.../stocks_1h_clean.csv",
  "featureSummary": {
    "missingPct": 0.02, "outlierEvents": 14, "alignedCalendars": true
  },
  "qualityFlags": {
  "latencyChecked": true,
  ...
  }
}
\end{verbatim}

\dashline

\textbf{Data Agent – BTC}\\
\textbf{Prompt:}\\
You are the Data Agent for BTC intraday data. Fetch, clean, deduplicate,
and timestamp-align minute-level bars. You may not reference future data.
Return training-window features only; do not compute P\&L or returns
beyond the provided lookback. BTC data typically comes from Polygon or
Binance public APIs.

\medskip
\textbf{Context:}\\
role=DataAgent, assetType=crypto, frequency="1m", lookbackDays=30

\medskip
\textbf{Input:}\\
{"symbols":["BTCUSDT"],\\
 "sourceHints":["polygon","binance"],\\
 "requiredFields":["open","high","low","close","volume"]}

\medskip
\textbf{Output (JSON):}
\begin{verbatim}
{
  "cleanedDataRef": "s3://.../btc_1m_clean.csv",
  "featureSummary": {
    "missingPct": 0.004, "dupTicksRemoved": 182, "timestampAligned": true
  },
  "qualityFlags": {
    "latencyChecked": true,
    ...
  }
}
\end{verbatim}
\end{tcolorbox}

\clearpage

\begin{tcolorbox}
\textbf{Alpha Agent – Stocks (Momentum)}\\
\textbf{Prompt:}\\
You are the Alpha Agent for a daily US equity momentum strategy. Propose
factor structures and signal formulations using only training-window feature
summaries and published academic or industry priors. Do not use any
evaluation-window data. Do not tune parameters based on Sharpe, IC,
t-statistics, or any performance metric from the evaluation horizon. All
numerical evaluation and backtesting are delegated to tools. Provide factor
intuition, mathematical form, and expected direction only.

\medskip
\textbf{Context:}\\
role=AlphaAgent, assetType=equity, horizonDays=5, family="momentum"

\medskip
\textbf{Input:}\\
{"featureSummary":{"returns\_20d":{"std":0.18},\\
"returns\_60d":{"std":0.30}},\\
"universe":["AAPL","MSFT","AMZN", ...],\\
"regimeHint":"calm"}

\medskip
\textbf{Output (JSON):}
\begin{verbatim}
{
  "factors": [
    {
      "name": "mom20_over_60",
      "family": "momentum",
      "type": "time_series",
      "horizonDays": 5,
      "lookbackDays": [20, 60],
      "formula": "returns_20d - returns_60d",
      "requiredInputs": ["returns_20d", "returns_60d"],
      "expectedDirection": "positive"
    },
    {
      "name": "cross_sectional_momentum",
      "family": "momentum",
      "type": "cross_sectional",
      "horizonDays": 5,
      "lookbackDays": [120],
      "formula": "rank(returns_120d)",
      "requiredInputs": ["returns_120d"],
      "expectedDirection": "positive"
    }
  ],
  "notes": {
    "usesTrainingWindowOnly": true,
    "noSharpeTuning": true,
    "leakageChecked": true
  }
}
\end{verbatim}

\end{tcolorbox}
\clearpage
\begin{tcolorbox}
\textbf{Alpha Agent – BTC (Microstructure)}\\
\textbf{Prompt:}\\
You are the Alpha Agent for BTC–USD intraday trading. Propose
microstructure-style factors (order-flow imbalance, realized volatility,
spread, volume spikes) using only training-window summaries and tool-provided
statistics. Do not use test labels, realized P\&L, or backtest metrics. Do not
search for the “best Sharpe” factor. Output factor definitions and expected
direction only.

\medskip
\textbf{Context:}\\
role=AlphaAgent, assetType=crypto, horizonMinutes=30,\\
family="microstructure"

\medskip
\textbf{Input (example):}
\begin{verbatim}
featureSummary: {
  "ret_5m": {"std": 0.45},
  "ret_30m": {"std": 0.80},
  "of_imbalance_1m": {"mean": 0.02},
  "spread_bp_1m": {"mean": 3.1},
  ...
}
regimeHint: "highVol"
\end{verbatim}

\medskip
\textbf{Output (JSON example):}
\begin{verbatim}
{
  "factors": [
    {
      "name": "short_term_momentum_5m",
      "family": "momentum",
      "type": "time_series",
      "horizonMinutes": 15,
      "lookbackMinutes": 5,
      "formula": "ret_5m",
      "requiredInputs": ["ret_5m"],
      "expectedDirection": "positive"
    },
    {
      "name": "order_flow_imbalance_5m",
      "family": "microstructure",
      "type": "time_series",
      "horizonMinutes": 15,
      "lookbackMinutes": 5,
      "formula": "buy_volume_5m - sell_volume_5m",
      "requiredInputs": ["buy_volume_5m", "sell_volume_5m"],
      "expectedDirection": "positive"
    }
  ],
  "notes": {
    "usesTrainingWindowOnly": true,
    "noSharpeTuning": true
  }
}
\end{verbatim}

\end{tcolorbox}

\clearpage

\begin{tcolorbox}
\textbf{Risk Agent – Stocks}\\
\textbf{Prompt:}\\
You are the Risk Agent for an equity portfolio. Interpret exposures and
volatility signals using tool-provided covariance matrices and factor models.
You may not access evaluation-window returns or future prices. You may not
modify alpha signals directly; only produce risk reports and binary gates.

\medskip
\textbf{Context:}\\
role=RiskAgent, assetType=equity, targetVol=0.15, maxLeverage=1.5

\medskip
\textbf{Input:}\\
{"alphaRef":"s3://.../us\_alpha.parquet",\\
 "sigmaRef":"s3://.../us\_Sigma.parquet",\\
 "exposureRef":"s3://.../us\_B.parquet",\\
 "constraints":{"sectorLimit":0.30,"singleNameLimit":0.05}}

\medskip
\textbf{Output (JSON):}
\begin{verbatim}
{
  "riskReport": {
    "forecastVol": 0.13,
    "grossLeverage": 1.20,
    "netBeta": 0.02,
    "sectorBreaches": []
  },
  "gates": {
    "vol_ok": true,
    "beta_ok": true,
    "sector_ok": true,
    "leakageChecked": true
  }
}
\end{verbatim}

\dashline

\textbf{Risk Agent – BTC}\\
\textbf{Prompt:}\\
You are the Risk Agent for BTC intraday strategies. Using realized-vol
and drawdown estimates from tools, check whether the proposed exposure
satisfies volatility, position-size, and intraday drawdown limits. You may
not compute P\&L or see evaluation-window returns.

\medskip
\textbf{Context:}\\
role=RiskAgent, assetType=crypto, targetVol=0.80, maxPositionPct=0.8

\medskip
\textbf{Input:}\\
{"positionPct":0.60,\\
 "realizedVolEstimate":0.70,\\
 "intradayDrawdownEstimate":0.18}

\medskip
\textbf{Output (JSON):}
\begin{verbatim}
{
  "riskReport": {
    "realizedVolEstimate": 0.70,
    "intradayDrawdownEstimate": 0.18
  },
  "gates": {
    "vol_ok": true,
    "dd_ok": true,
    "position_ok": true,
    "leakageChecked": true
  }
}
\end{verbatim}
\end{tcolorbox}

\clearpage

\begin{tcolorbox}
\textbf{Portfolio Agent – Stocks}\\
\textbf{Prompt:}\\
You are the Portfolio Agent for US equities. Combine admissible alpha signals
and risk diagnostics into portfolio weights under the constraints provided by
the orchestration plan. Do not perform backtests or access realized returns.
All numerical optimization is delegated to tools (e.g., convex solvers). You
only specify the optimization problem and return the resulting weight vector
and metadata.

\medskip
\textbf{Context:}\\
role=PortfolioAgent, assetType=equity, turnoverLimit=0.20, tcModel="linear"

\medskip
\textbf{Input:}\\
{"alphaRef":"s3://.../us\_alpha.parquet",\\
 "riskGates":{"vol\_ok":true,"beta\_ok":true},\\
 "capital":100000}

\medskip
\textbf{Output (JSON):}
\begin{verbatim}
{
  "weights": {
    "AAPL": 0.12,
    "MSFT": 0.10,
    "GOOGL": 0.08,
    "others": "..."
  },
  "exposure": {
    "gross": 1.05,
    "net": 0.15
  },
  "notes": { 
  "turnoverRespected": true, "riskGatesPassed": true
  }
}
\end{verbatim}

\dashline

\textbf{Portfolio Agent – BTC}\\
\textbf{Prompt:}\\
You are the Portfolio Agent for BTC intraday trading. Adjust BTC exposure
based on admissible microstructure signals and risk gates. No returns or P\&L
may be accessed. Output a single BTC weight or position percentage.

\medskip
\textbf{Context:}\\
role=PortfolioAgent, assetType=crypto, turnoverLimit=0.50

\medskip
\textbf{Input:}\\
{"signalRef":"s3://.../btc\_signal.parquet",\\
 "riskGates":{"vol\_ok":true,"dd\_ok":true},\\
 "capital":100000}

\medskip
\textbf{Output (JSON):}
\begin{verbatim}
{
  "positionPct": 0.55,
  "exposure": {
    "gross": 0.55,
    "net": 0.55
  },
  "notes": {
    "turnoverRespected": true, "riskGatesPassed": true, ...
  }
}
\end{verbatim}
\end{tcolorbox}





 





\clearpage

\begin{tcolorbox}
\textbf{Backtest Agent – Stocks}\\
\textbf{Prompt:}\\
You are the Backtest and Evaluation Agent for a daily equity strategy. Using
stored signals and executed orders, compute portfolio-level metrics over the
evaluation window. You are the only component that touches
evaluation-window returns. Do not expose per-timestamp labels or raw P\&L to
any LLM agent; return only aggregated metrics and diagnostics.

\medskip
\textbf{Context:}\\
role=BacktestAgent, assetType=equity, evalWindow="2024-05-01:2024-12-31"

\medskip
\textbf{Input:}\\
{"ordersRef":"s3://.../orders.parquet",\\
 "priceRef":"s3://.../prices.parquet",\\
 "metrics":["vol","sharpe","maxdd","turnover"]}

\medskip
\textbf{Output (JSON):}
\begin{verbatim}
{
  "metrics": {
    "vol": 0.1183,
    "sharpe": 2.63,
    "maxDrawdown": -0.0359,
    "turnover": 0.21
  },
  "notes": {
    "noTimestampPnL": true,
    "leakageChecked": true
  }
}
\end{verbatim}

\dashline

\textbf{Backtest Agent – BTC}\\
\textbf{Prompt:}\\
You are the Backtest Agent for BTC minute-level strategies. Evaluate the
strategy over the given intraday evaluation window. No timestamp-level P\&L
or trade-by-trade returns may be exposed; only aggregated metrics are allowed.

\medskip
\textbf{Context:}\\
role=BacktestAgent, assetType=crypto, evalWindow="2025-07-27:2025-08-13"

\medskip
\textbf{Input:}\\
{"ordersRef":"s3://.../btc\_orders.parquet",\\
 "priceRef":"s3://.../btc\_prices.parquet",\\
 "metrics":["vol","sharpe","maxdd"]}

\medskip
\textbf{Output (JSON):}
\begin{verbatim}
{
  "metrics": {
    "vol": 0.412,
    "sharpe": 0.174,
    "maxDrawdown": -0.0091
  },
  "notes": {
    "noTimestampPnL": true,
    "leakageChecked": true
  }
}
\end{verbatim}
\end{tcolorbox}

\clearpage

\section{Context Protocols}

The orchestration system uses structured context messages to pass information between agents. All contexts are serialized as JSON and follow the schema
\begin{equation}
    C = \{ \text{task\_id}, \text{agent\_role}, \text{run\_mode},
    \text{time\_window}, \text{universe}, \text{inputs},
    \text{tool\_outputs}, \text{diagnostics}, \text{uuid} \}.
\end{equation}
Here \texttt{run\_mode} $\in \{\text{train},\text{test},\text{live}\}$ and \texttt{time\_window} records the lookback and horizon for the current step.

Each context message excludes:
\begin{itemize}
    \item any raw price or return series from the test period;
    \item any labels or targets from future timestamps;
    \item any optimization objective that is tied directly to the evaluation window (for example, Sharpe ratio on the test set).
\end{itemize}

Numerical arrays are not sent directly. Instead, they are stored in data files or tables and referred to by identifiers (for example, data paths or dataset IDs). Only the Backtest Agent is allowed to load data that belong to the evaluation window. Contexts are written through the Memory Agent together with their \texttt{uuid} so that runs can be reproduced, checked, and replayed later.

\section{Memory Integration and UUID Protocols}

The Memory Agent stores long-term states indexed by deterministic UUIDs. Each UUID is computed as a cryptographic hash of the agent role, task description, parameter configuration, and timestamp:
\begin{equation}
    \text{UUID} = \text{SHA256}(\text{role} \Vert \text{task}
    \Vert \text{params} \Vert \text{time}).
\end{equation}
This connects each memory record to a specific prompt and run setup, which helps with reproducibility and safety.

The UUID design supports:
\begin{itemize}
    \item \emph{immutability}: once written, memory entries are referred to only by their hash ID;
    \item \emph{identity matching}: runs with the same configuration can look up compatible past states across orchestration cycles;
    \item \emph{safe retrieval}: downstream agents query by UUID and receive only summarized metadata, not raw test-set values;
    \item \emph{isolation}: training and evaluation memories use separate UUID namespaces to avoid mixing information.
\end{itemize}

A typical stored memory entry has the form
\begin{equation}
    M = \{ \text{uuid}, \text{agent\_role}, \text{plan\_step},
    \text{features\_hash}, \text{metrics\_summary}, \text{timestamp} \},
\end{equation}
where \texttt{features\_hash} is a non-invertible checksum of the input data and \texttt{metrics\_summary} contains only aggregated statistics (for example, average IC by bucket or gate pass rates). Memory entries never store raw prices, raw returns, or full P\&L paths, so they cannot be used to reconstruct evaluation-period labels.

\section{Leakage Prevention Summary}

Across all agents, prompts and context protocols enforce a clear separation between LLM-based reasoning and numerical computation. LLM agents never receive evaluation-window returns, prices, or labels. Optimization and backtesting are implemented as deterministic tools behind the orchestration layer, and their outputs are filtered before any feedback is used in LLM prompts. UUID-based memory records make it easy to tell which run a record belongs to and to repeat the same run later, while storing only simple summaries that cannot be turned back into raw test data. These design choices lower the chance of data leakage and help keep our agentic trading experiments valid under walk-forward evaluation.

\end{document}